\documentclass[aps,prl,preprint,superscriptaddress, showpacs, floatfix]{revtex4-1}

\usepackage[utf8]{inputenc}

\usepackage{amsmath}
\usepackage{amssymb}
\usepackage{latexsym}

\usepackage{ifpdf}

\ifpdf
\usepackage[pdftex]{graphicx}
\else
\usepackage{graphicx}
\fi

\ifpdf
\DeclareGraphicsExtensions{.pdf, .jpg, .tif}
\else
\DeclareGraphicsExtensions{.eps, .jpg}
\fi

\begin{document}


\title{The Casimir force on a surface with shallow nanoscale corrugations: Geometry and finite conductivity effects}


\author{Y. Bao}
\affiliation{Department of Physics, University of Florida, Gainesville, Florida 32611}
\author{R. Guérout}
\author{J. Lussange}
\author{A. Lambrecht}
\affiliation{Laboratoire Kastler–Brossel, CNRS, ENS, Université Pierre et Marie Curie case 74, Campus Jussieu, F-75252 Paris Cedex 05, France}
\author{R. A. Cirelli}
\author{F. Klemens}
\author{W. M. Mansfield}
\author{C. S. Pai}
\affiliation{Bell Laboratories, Alcatel-Lucent, Murray Hill, NJ 07974, USA}
\author{H. B. Chan}
\affiliation{Department of Physics, the Hong Kong University of Science and Technology, Hong Kong, China}
\email{hochan@ust.hk}

\begin{abstract}
We measure the Casimir force between a gold sphere and a silicon plate with nanoscale, rectangular corrugations with depth comparable to the separation between the surfaces. In the proximity force approximation (PFA), both the top and bottom surfaces of the corrugations contribute to the force, leading to a distance dependence that is distinct from a flat surface. The measured Casimir force is found to deviate from the PFA by up to $15\%$, in good agreement with calculations based on scattering theory that includes both geometry effects and the optical properties of the material.
\end{abstract}


\maketitle
The Casimir force between two neutral conductors arises from the change of the zero point energy associated with quantum fluctuation of the electromagnetic field in the presence of boundaries. Between two parallel plates, the Casimir force is attractive and its magnitude increases rapidly as the separation decreases. In recent years, the Casimir force has received significant attention, from fundamental interests to possible applications in micro and nano-electromechanical systems \cite{Lamoreaux1997, Mohideen1998, Roy1999, Chan2001, Bressi2002, Decca2003,Chen2006, Chan2008, Man2009, Jourdan2009, Munday2009, Chiu2010}. For instance, fundamental questions on how to account for the temperature corrections to the Casimir force remain a controversial topic \cite{Klimchitskaya2009}. At the same time, there has been much progress in the control of the Casimir force by modifying the optical properties of the interacting surfaces, such as using dissimilar metals \cite{Decca2003}, replacing one surface with semiconductors with different carrier concentrations \cite{Chen2006}, and inserting fluid into the gap between the surfaces \cite{Munday2009}. In addition, a number of efforts aim at generating repulsive Casimir forces with a vacuum gap using metamaterials \cite{rosa2008, Zhao2009}. 

Apart from the optical properties of the material, the Casimir force depends on the shape of the interacting objects in non-trivial ways. For small deviations from the planer geometry, the Casimir force can be estimated by the proximity force approximation (PFA) \cite{Buscher2004}. In the common experimental configuration of sphere and plate, the PFA works well provided that the separation is much smaller than the radius of the sphere. However, the PFA breaks down for other geometries. Theoretical analysis indicates that for a thin conducting spherical shell \cite{Boyer1968} or a rectangular box with a certain aspect ratio \cite{Maclay2000}, the Casimir energy has opposite sign to parallel plates, opening the possibility of generating repulsive Casimir forces. Advanced theoretical approaches are now capable of calculating the Casimir force between structures of arbitrary shapes \cite{Lambrecht2006, Rahi2009, Reid2009}. These approaches are not limited to perfectly conducting objects, but can also take into account the optical properties of the material. Experimentally, revealing the strong geometry dependence of the Casimir force involves introducing deformations on a planar surface. The first such attempt was performed by Roy and Mohideen, who measured the Casimir force on surfaces with small sinusoidal corrugations \cite{Roy1999}. Subsequently, the lateral Casimir force in similar structures has been demonstrated by the same team to deviate from the PFA \cite{Chiu2010}. Recently, we measured the Casimir force on a surface with an array of high aspect ratio trenches \cite{Chan2008}. Deviation of up to $20\% $ from PFA is observed. While this experiment provides evidence for the non-trivial boundary dependence of the Casimir force, the measured results are smaller than the predicted values for perfect metallic structures of the same geometry \cite{Buscher2004}. It becomes apparent that meaningful comparison of experimental results to theory would require both geometry effects and finite conductivity of the material to be included. 

In our previous experiment \cite{Chan2008}, we considered the Casimir force between a surface with an array of deep rectangular trenches and another flat surface on top. The trench array is assumed to have solid volume fraction equal to $p$. In the PFA picture, the total interaction is a sum of two contributions: (1) the interaction between a fraction $p$ of the flat surface and the top surface of the trench array separated by distance $z$; and (2) the interaction between a fraction of $(1-p)$ of the flat surface and the bottom of the trench array at distance $z+a$, where $a$ is the depth of the trenches. The second contribution is negligible for such deep trenches because the Casimir force at this separation ($z+a>1\:\mu $m) is too small to be detected in our measurement setup. Therefore, under the PFA, the force on the trench array is practically identical to the force between two parallel flat surfaces at separation $z$ multiplied by a constant factor $p$. In other words, for the deep trenches, the distance dependence of the force under the PFA is the same as a flat surface.

In this Letter, we report measurements of the Casimir force between a gold sphere and a silicon plate with nanoscale, rectangular corrugations with depth comparable to the separation between the surfaces. In the PFA, both the top and bottom surfaces of the corrugations contribute to the force, yielding a distance dependence that is distinct from a flat surface. The measured Casimir force is found to deviate from the PFA by about $10\% $. 
We present calculations based on scattering theory that includes the finite conductivity of silicon, yielding good agreement with measurement. Our results demonstrate that for surfaces with nanoscale deformations, the Casimir force depends on a profound interplay between geometry effects and material properties.

\begin{figure}%
\includegraphics[angle=0, width = 3.2in]{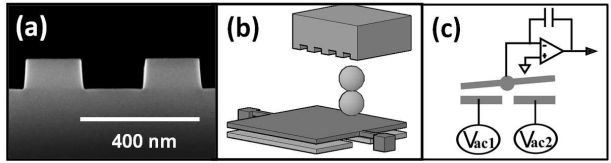} 
\caption{\label{fig:fig1} (a) Scanning electron micrograph of the cross section view of the trench array. (b) Schematic of the experimental setup (not to scale). (c) Measurement scheme with electrical connections. $V_{ac1}$ and $V_{ac2}$ are the excitation voltages applied to the bottom of the electrodes.}
\end{figure}

Figure \ref{fig:fig1}(a) shows a scanning electron micrograph of the cross section of the trench array with periodicity of 400$\:$nm. We fabricate the trenches by dry etching into a highly p-doped silicon wafer with a lithographically defined silicon oxide pattern as the etch mask. In the reactive ion etching step, an inductively-coupled plasma of $SF_6$ and $Ar$ was used without any passivation gas. The reactant flow rate, pressure and bias were optimized to yield a smooth and flat bottom surface so that its contribution to the PFA can be easily determined. Such a recipe, however, produced a sidewall at $94.6\,^\circ$ to the top surface, close to but not exactly vertical. After etching, the oxide mask is removed using hydrofluoric acid (HF). Another sample, consisting of a flat surface with no corrugations, is also prepared. Both samples are fabricated from the same wafer to ensure that the optical properties of the silicon are identical. 

Accurate determination of the dimensions of the trench array is crucial in the electrostatic force and the Casimir force calculations. Ten cross section views [similar to Fig. \ref{fig:fig1}(a)] at different positions of the trench array are taken using a scanning electron microscope (SEM). The lengths of the top surface and the bottom surface in one period are measured to be $l_1=185.3\:$nm and $l_2=199.1\:$nm respectively. An atomic force microscope is used to obtain the depth of the trenches. The average of one set of ten scans of $2\:\mu $m square and another set of $1\:\mu$m square at different locations gives $t = 98\pm0.7\:$nm. This depth is chosen to be smaller than the typical separation between the two interacting bodies, so that the force from the bottom surface is not negligible if the PFA is assumed to be valid:
\begin{eqnarray}
\label{eq:eq1}
F_{PFA} &=& (1/\lambda)\int_{0}^{\lambda} F_{flat}(z(x)) dx \nonumber\\
		&=& p_1F_{flat}(z)+p_2F_{flat}(z+t)+2\int^{p_3}_{0}F_{flat}(z+tx/p_3)dx,
\end{eqnarray}
where $F_{flat}$ is the force on a flat surface made of the same material, $p_1 = l_1/\lambda$, $p_2 = l_2/\lambda$ and $p_3 = (1- p_1- p_2)/2$. In Eq. (\ref{eq:eq1}), the first two terms represent the contributions of the top and bottom surfaces respectively, accounting for $\sim 97 \%$ of the force under the PFA. The third term introduces a small modification originating from the sidewalls that are not perfectly vertical. While deriving the force on such corrugated structures using the PFA is rather straight forward, the actual Casimir force is expected to deviate from the PFA due to its non-trivial dependence on the geometry of the interacting objects. Since such deviations increase with the ratio $z/\lambda$ \cite{Buscher2004}, the corrugated sample is chosen to have the smallest $\lambda$ that can be reproducibly fabricated with our lithography and etching tools. Calculations of the Casimir force on this exact geometry using scattering theories will be presented later. 

Figure \ref{fig:fig1}(b) shows a schematic (not to scale) of a micromechanical oscillator that measures the force gradient between the corrugated surface and a spherical surface. The oscillator is made of a $3.5\:\mu$m thick, $500\:\mu$m square heavily doped polysilicon plate suspended by two torsional rods. Underneath the oscillator's top plate, there are two fixed electrodes. 
Torsional oscillations of the top plate are electrostatically excited when a small ac voltage close to the resonant frequency of the oscillator ($f_0 = 1783\:$Hz and quality factor 32,000) is applied to one electrode. 
Motion of the top plate is detected by the capacitance change between the top plate and the electrodes using additional ac voltages at amplitude of 100$\:$mV and frequency of 102$\:$kHz. 
Two glass spheres, each with radius $R=50\:$$\mu$m, 
are coated with a layer of gold with thickness of 4000$\:$A. 
They are stacked and attached onto one side of the top plate using conductive epoxy at a distance of $b=210\:\mu$m from the rotation axis. 

Preparation of the silicon surfaces involves a number of important steps. First, the native oxide on the surfaces of the silicon samples was removed by HF. This procedure also passivates the silicon surface so that oxide does not re-form in ambient pressure for a few hours \cite{Chen2006}. To eliminate residual water on the corrugations, the silicon chip was baked at $120\,^\circ$C for 15 minutes. Afterwards, the silicon sample is positioned face down at a few $\mu$m from the top of the spheres. The chamber is then immediately evacuated to a base pressure of $10^{-6}$ torr by dry pumps.

A closed-loop piezoelectric actuator controls the distance between the silicon sample and the sphere. The distance $z$ is given by $z=z_0-z_{piezo}-b\theta$, where $z_0$ is the initial gap between two surfaces, $z_{piezo}$ is the piezo extension and $b\theta$ is a correction term to account for the tilting angle $\theta$ of the top plate. A phase locked loop is used to track the frequency shift of the oscillator as the sphere approaches the silicon sample. At small oscillations where nonlinear effects can be neglected, the shift in the resonant frequency is proportional to the force gradient
\begin{equation}
\label{eq:eq2}
\Delta f=C\frac{\partial F}{\partial z}
\end{equation}
where $C = -b^2/8\pi ^2If_0$ and $I$ is the moment of inertia of the top plate together with the two spheres. The oscillation amplitude of the oscillator is reduced as z decreases to avoid the oscillation from becoming nonlinear. 

\begin{figure}%
\includegraphics[angle=0, width = 2in]{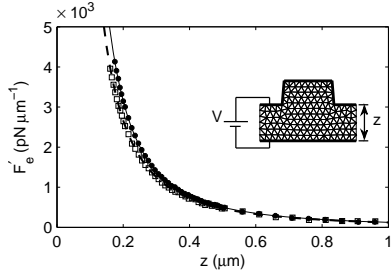} 
\caption{\label{fig:fig2}The electrostatic force gradient as a function of distance for $V= V_0+300$ mV on the flat silicon surface (solid circles) and corrugated silicon structure (hollow squares). The solid line is a fit using Eq. (3) for a flat surface and the dash line is a fit using the numerical calculations for the corrugated structure. Inset: Meshing of the gap between the two surfaces to solve the Poisson equation in 2D ($z=150\:$nm). The number of triangles is 40 times larger in the actual calculation.}

\end{figure}

We apply electrostatic forces to calibrate the constant $C$ and the initial distance between the surfaces $z_0$. The electrostatic force between the grounded gold sphere and the flat plate at voltage $V$ is given by:
\begin{equation}
\label{eq:eq3}
F_e = 2\pi \epsilon_0(V-V_0)^{2}\sum _{n=1}^{\infty}\frac{[\coth (\alpha)-n\coth (n\alpha)]}{\sinh(n\alpha)},
\end{equation}
where $\epsilon_0$ is the permittivity of vacuum, $\alpha = \cosh^{-1}(1+d/R)$ and $d$ is the separation between the sphere and the plate. The residual voltage $V_0$ is measured to be $-0.499\:$V by finding the voltage at which the frequency shift $\Delta f$ attains minimum at a fixed distance. $V_0$ is found to change by less than 3$\:$mV for $z$ ranging from 100$\:$nm to 600$\:$nm. In Fig. \ref{fig:fig2}, the solid circles represent the measured electrostatic force gradient on the flat silicon sample at $V-V_0 = 300\:$mV and the solid line is a fit using Eqs. (\ref{eq:eq2}) and (\ref{eq:eq3}) after subtracting the contribution of the Casimir force (the measurement of which is described later). $C$ is determined to be $614\pm3\:$m$\:N^{-1}\:s^{-1}$ by averaging six sets of data with $V-V_0$ between 245$\:$mV and 300$\:$mV. For the corrugated silicon sample, the calibration procedure is similar. However, since there is no analytic expression for the electrostatic force, it is necessary to solve Poisson's equation in 2D numerically. The boundary conditions, as shown in the inset of Fig. \ref{fig:fig2}, are set by maintaining a fixed potential between the trench array and a flat surface, with periodic boundary conditions applied to one period of the array. Then, the potential distribution is calculated using finite element analysis, with the confined area divided into $N>10,000$ triangles. Since $R>>z$, the proximity force approximation $F_{s,grat} = 2 \pi R E_{f,grat}$ is used to obtain the force $F_{s,grat}$ between a sphere and a corrugated surface, where $E_{f,grat}$ is the electrostatic energy per unit area between a flat surface and a corrugated surface. To ensure the convergence of the numerical calculation, we checked that the calculated force varies by less than $0.1\%$ even when $N$ is doubled. 

\begin{figure}%
\includegraphics[angle=0, width = 3.2in]{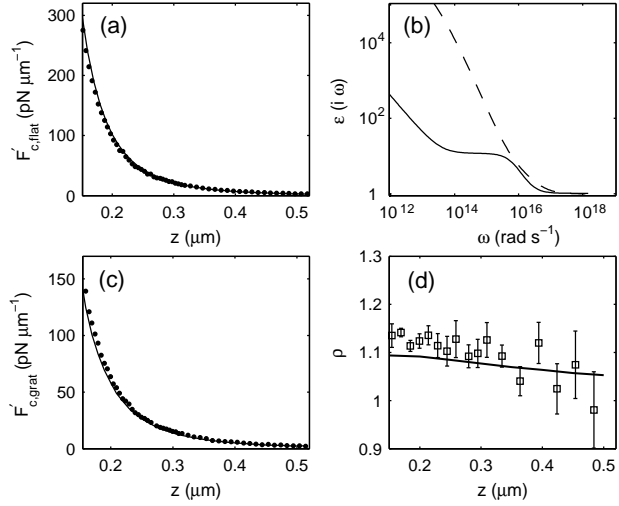} 
\caption{\label{fig:fig3}(a) Measured Casimir force gradient between the gold sphere and the flat silicon surface $F^{\prime}_{c,flat}$. The solid line represents the theoretical calculation including finite conductivity and surface roughness corrections. (b) Dielectric functions evaluated at imaginary frequencies for doped silicon (plain line) and gold (dashed line). (c) Measured Casimir force gradient on corrugated silicon structure. The line represents the force gradient expected from the PFA. (d) The squares are the ratio $\rho$ of the measured Casimir force gradient to the force gradient expected from the PFA. The solid line plots the theoretical values including both geometry and finite conductivity effects. 
}
\end{figure}

Next, the Casimir force gradient $F^{\prime}_{c,flat}$ on the flat silicon surface is measured by setting $V$ equal to $V_0$. In Fig. \ref{fig:fig3}(a), the circles are the measured data and the solid line represents the theoretical values. To account for the finite conductivity of the materials, the dielectric functions evaluated at imaginary frequencies $\epsilon(i\omega)$ are used in Lifshitz's formula. For gold, we use optical data extrapolated at low frequencies by the Drude model $\epsilon_g(i\omega)=1+\frac{\omega_{p,g}^2}{\omega(\omega+\gamma_g)}$ with a plasma frequency $\omega_{p,g}=9\:$eV and a relaxation rate $\gamma_g=35\:$meV. For silicon, the Drude-Lorentz model is used: $\epsilon_{si}(i\omega)=\epsilon_{i}(i\omega)+\frac{\omega_{p,si}^2}{\omega(\omega+\gamma_{si})}$. $\epsilon_{i}(i\omega)$ is the dielectric function for intrinsic silicon, taken from Ref. \cite{Lambrecht2007}. The plasma frequency $\omega_{p,si}$ ($1.36\times10^{14}$ rad.s$^{-1}$) and the relaxation rate $\gamma_{si}$ ($4.75\times10^{13}\:$rad.s$^{-1}$) are interpolated from the data in Ref. \cite{Duraffourg2006} for a carrier density of $2\times10^{18}\:$cm$^{-3}$ determined from the dc conductivity of the wafer. Figure \ref{fig:fig3} (b) shows the dielectric functions used for doped silicon and gold. The force calculated by Lifshitz's formula is further modified by the roughness correction using the geometrical averaging method \cite{Chen2006-2}. The contribution to the roughness correction originates mainly from the gold surface ($\sim 4\:$nm rms) rather than the silicon wafer ($\sim0.6\:$nm rms). 

The Casimir force gradient $F^{\prime}_{c,grat}$ between the same gold sphere and the corrugated silicon sample is then measured  and plotted as circles in Fig. \ref{fig:fig3}(c). Comparison to the PFA is performed by evaluating Eq. (\ref{eq:eq1}) with the measured Casimir force on the flat silicon surface. As described earlier, the force gradient on the corrugations under the PFA, $F^{\prime}_{c,PFA}$, is the sum of the force on the top and bottom surfaces, with a small contribution from the slightly slanted sidewalls. The deviations of the measured Casimir force from the PFA arise due to the strong geometry dependence of the Casimir force. For a more quantitative analysis of the deviation, the ratio $\rho  = F^{\prime}_{c,grat} / F^{\prime}_{c,PFA}$ is plotted in Fig. \ref{fig:fig3}(d). The measured $F^{\prime}_{c,grat}$ clearly exceeds $F^{\prime}_{c,PFA}$, by up to $15\%$.

We perform exact calculations for the Casimir force $F_{c,grat}(z)$ per unit area between a flat gold plate and the corrugated silicon surface, taking into account the non-specular reflections introduced by the grating structure. Then, we use the PFA to relate the sphere-plane and the plane-plane geometries according to $F^{\prime}_{c,grat}=2\pi R F_{c,flat}$. The theory for calculating the Casimir energy based on scattering theory \cite{Lambrecht2006} for structures involving gratings has been presented elsewhere \cite{Lambrecht2008} and will be only briefly summarized. The zero temperature Casimir force per unit area between two reflecting objects separated by a distance $z$ is 
\begin{equation}
  \label{eq:CasForce}
  F=-\hbar\iiint\text{tr}\left(\left(\mathbf{1}-\boldsymbol{\mathcal{M}}\right)^{-1}\partial_{z}\boldsymbol{\mathcal{M}}\right)\,\mathrm{d}^{2}\mathbf{k}_{\perp}\,\mathrm{d} \xi
\end{equation}
where $\mathbf{k}_{\perp}$ gather the components of the wave vector in the plane of the objects and $\xi=i \omega$ is the Wick-rotated imaginary frequency. $\boldsymbol{\mathcal{M}}$ is the open-loop function $\boldsymbol{\mathcal{M}}=\mathbf{R_{1}}(\xi)e^{-\boldsymbol{\kappa}z}\mathbf{R_{2}}(\xi)e^{-\boldsymbol{\kappa}z}$ with $\mathbf{R_{1}}$ and $\mathbf{R_{2}}$ the reflection operators for the two objects and $\boldsymbol{\kappa}=\sqrt{\xi^{2}/c^{2}+\mathbf{k}_{\perp}^{2}}$. For planar objects, the reflection operators are diagonal in the plane wave basis and collect the appropriate Fresnel coefficients. For gratings, this does not hold anymore. The reflection operators are not diagonal as they mix different polarizations and account for non specular reflections. Therefore, in general the matrices $\boldsymbol{R_{i}}$ and $e^{-\boldsymbol{\kappa}z}$ do not commute and we write $-\partial_{z}\boldsymbol{\mathcal{M}}=\mathbf{R_{1}}(\xi)\boldsymbol{\kappa}e^{-\boldsymbol{\kappa}z}\mathbf{R_{2}}(\xi)e^{-\boldsymbol{\kappa}z}+\mathbf{R_{1}}(\xi)e^{-\boldsymbol{\kappa}z}\mathbf{R_{2}}(\xi)\boldsymbol{\kappa}e^{-\boldsymbol{\kappa}z}$. The results of the exact calculation, normalized by the PFA, are plotted as the solid line in Fig. \ref{fig:fig3}(d), yielding good agreement with measurements. 

Our results demonstrate that it is possible to both calculate and measure the Casimir force in nanostructured surfaces of unconventional shapes with high accuracy. The interplay between finite conductivity and geometry effects holds promise as an important tool to control the Casimir force between mechanical components at close proximity.

Y.B. and H.B.C. are supported by DOE Grant No. DE-FG02-05ER46247 and NSF Grant No. DMR-0645448. R.G., J.L. and A.L. are supported by the European Science Foundation (ESF) within the activity “New Trends
and Applications of the Casimir Effect” (www.casimir-network.com) and by the French National Research Agency (ANR) through grant No ANR-06-NANO-062 - MONACO project.

\end{document}